\documentclass[aps,pra,reprint,superscriptaddress, floatfix]{revtex4-2}

\usepackage{graphicx}
\usepackage{physics}
\usepackage{amsmath}
\usepackage{amssymb}
\usepackage{bm}
\usepackage{xcolor}
\usepackage[most]{tcolorbox}
\usepackage{soul}
\sethlcolor{yellow}
\usepackage{hyperref}

\begin{document}

\title{Hardware Demonstration of Neural-Surrogated Quantum-Enhanced MCMC for Large-Scale Constrained Boltzmann Sampling}

\author{Yuya Kawamata}
\email[Email: ]{u597035b@ecs.osaka-u.ac.jp}
\affiliation{Graduate School of Engineering Science, The University of Osaka,
1-3 Machikaneyama, Toyonaka, Osaka 560-8531, Japan}
\author{Keisuke Fujii}
\affiliation{Graduate School of Engineering Science, The University of Osaka,
1-3 Machikaneyama, Toyonaka, Osaka 560-8531, Japan}
\affiliation{Center for Quantum Information and Quantum Biology, The University of Osaka, 1-2 Machikaneyama, Toyonaka, 560-0043, Japan.}
\affiliation{RIKEN Center for Quantum Computing (RQC), Hirosawa 2-1, Wako, Saitama 351-0198, Japan}
\affiliation{Graduate School of Informatics, Kyoto University, Sakyo-ku, Kyoto, 606-8501, Japan}

\date{\today}

\begin{abstract}
Neural-surrogated quantum-enhanced Markov chain Monte Carlo (QeMCMC) uses neural networks trained on samples from the quantum approximate optimization algorithm (QAOA) to accelerate Boltzmann sampling.
A divide-and-conquer strategy extends this approach to larger systems, but independently optimizing QAOA parameters for each subproblem is costly.
In this paper, we develop a hardware-efficient implementation of neural-surrogated QeMCMC and demonstrate it using an IBM Quantum Heron r2 processor.
Experiments with $N=144$ binary variables reveal faster autocorrelation decay for appropriate subproblem sizes while exposing challenges in neural-surrogate learning for larger blocks.
We then extend the demonstration to a system with $N=1064$, where the autocorrelation decay rate per MCMC step is approximately $5.2$ and $12.3$ times those of global and local Kawasaki dynamics, respectively.
These results demonstrate faster autocorrelation decay in a system exceeding one thousand variables using samples generated on quantum hardware.
\end{abstract}

\maketitle

\section{Introduction}
Quantum computers have the potential to outperform classical computers for certain computational problems and are expected to find applications in optimization, quantum chemistry, machine learning, statistical physics, and other fields \cite{shor1994algorithms,grover1997quantum,abrams1999quantum,harrow2009quantum}.
However, currently available quantum devices are noisy intermediate-scale quantum (NISQ) devices \cite{preskill2018quantum}, and it remains difficult to run large-scale quantum algorithms with full error correction.
Therefore, variational quantum algorithms (VQAs) have been widely studied as a framework for using NISQ devices \cite{cerezo2021variational}.
A representative example is the quantum approximate optimization algorithm (QAOA) \cite{farhi2014quantum}, which uses relatively shallow parameterized quantum circuits to minimize a cost function.
However, scaling VQAs to larger problems remains challenging because of optimization difficulties such as barren plateaus \cite{mcclean2018barren}, the large number of measurements required to estimate expectation values, and the increasing effect of hardware noise with circuit size \cite{gonthier2022measurements}.

One promising way to exploit current quantum devices is to use samples generated by quantum circuits in combination with classical computation \cite{arute2019quantum,zhong2020quantum,madsen2022quantum}.
In particular, quantum-enhanced Markov chain Monte Carlo (QeMCMC) uses quantum-generated samples as proposal distributions and can provide nonlocal updates that accelerate MCMC mixing \cite{layden2023quantum,nakano2024markov}.
Previous work proposed distilling QAOA sample distributions into masked autoencoders for distribution estimation (MADEs) \cite{germain2015made,uria2016neural,larochelle2011neural} and using the learned distributions as Metropolis--Hastings proposals for Boltzmann sampling \cite{nakano2025neural}.
Furthermore, a divide-and-conquer strategy extends this approach to systems larger than the available quantum processor by applying QAOA and surrogate learning to individual subproblems \cite{kawamata2026divide}.
However, implementing this approach at large scale on real quantum hardware requires substantial quantum resources.
In particular, optimizing QAOA parameters separately for many subproblems requires repeated expectation-value evaluations on the quantum processor, which leads to a large number of quantum-circuit executions as the number of subproblems increases \cite{gonthier2022measurements}.

In this work, we develop a hardware-efficient implementation of neural-surrogated QeMCMC and demonstrate constrained Boltzmann sampling for systems with up to $N=1064$ binary variables using samples generated on an IBM Quantum Heron r2 processor.
Specifically, we construct subproblems with similar graph topologies and employ linear-ramp QAOA \cite{montanez2025linear}, which uses only four variational parameters while retaining high transferability across problem instances and sizes.
We optimize these four parameters only for one representative subproblem and reuse them for the remaining subproblems, exploiting the transferability of QAOA parameters between problems with similar graph structures and optimization landscapes \cite{galda2021transferability,galda2023similarity,montanez2025linear}.
The remaining subproblems therefore require only sampling with the transferred parameters, eliminating separate parameter optimization and substantially reducing the number of quantum-circuit executions on hardware.
Although one might expect parameter transfer to degrade performance, we found that it still accelerates MCMC convergence, as demonstrated by the numerical experiments mentioned later.

To evaluate the proposed method, we investigate its performance for $N=144$ binary variables by varying the subproblem size using samples generated on an IBM Quantum Heron r2 processor with $156$ qubits, and then demonstrate that it can be extended to a larger system with $N=1064$ binary variables.
More precisely, in the first experiment, we study constrained Boltzmann sampling with $N=144$ binary variables, using subproblem sizes $b=2, 12, 30, 49, 82,$ and $144$.
To reduce routing overhead in QAOA circuits, we construct QUBO problems whose interaction graphs match the connectivity of the quantum processor \cite{pelofske2023quantum,montanez2025evaluating,bucher2026constrained}.
We found that the autocorrelation decay rate increases with subproblem size up to a certain point but then decreases, which contrasts with previous classical simulations that showed continued improvement for sizes up to $b=16$ \cite{kawamata2026divide}.
Based on this finding, we construct a larger QUBO problem with $N=1064$ binary variables using the same connectivity pattern and demonstrate constrained Boltzmann sampling with $b=49$, a subproblem size that yielded fast autocorrelation decay in the first experiment.
The resulting autocorrelation decay rate per MCMC step is approximately $5.2$ and $12.3$ times those of nonlocal and nearest-neighbor pair-flip dynamics, respectively.
These results show that the improvement over the classical baselines is maintained when the full system size exceeds one thousand variables using samples from quantum hardware.

The remainder of this paper is organized as follows.
Sections~II and III introduce the sampling framework and its hardware-efficient implementation.
Sections~IV and V present the $N=144$ experiments and the $N=1064$ demonstration, respectively.
Section~VI summarizes the results and outlines future directions.

\section{Preliminaries}
\label{sec:preliminaries}

\subsection{Markov Chain Monte Carlo (MCMC)}

Markov chain Monte Carlo (MCMC) is a sampling method that constructs a Markov chain whose stationary distribution is a target probability distribution from which direct sampling is difficult.
In this study, we consider an energy function $E(\bm{x})$ defined for binary variables
$\bm{x}=(x_1,\ldots,x_N)\in\{0,1\}^{N}$.
Let $G=(V,\mathcal{E})$ denote the interaction graph with the vertices $N=|V|$.
The energy function is defined as
\begin{align}
E(\bm{x})
=
\sum_{(i,j)\in\mathcal{E}} Q_{ij}x_i x_j
+
\sum_{i\in V} q_i x_i.
\end{align}
Here, $Q_{ij}$ and $q_i$ denote the quadratic and linear coefficients, respectively.
In particular, we focus on Boltzmann sampling under the fixed Hamming-weight constraint
\begin{align}
\sum_{i=1}^{N} x_i
=
K.
\end{align}
The feasible set is defined as
\begin{align}
\Omega_K
=
\left\{
\bm{x}\in\{0,1\}^{N}
\,\middle|\,
\sum_{i=1}^{N}x_i=K
\right\}.
\end{align}
The corresponding constrained Boltzmann distribution is
\begin{align}
\label{eq:constrained-boltzmann}
\pi_K(\bm{x})
=
\frac{1}{Z_K}
\exp\left[-\beta_{\pi}E(\bm{x})\right],
\qquad
\bm{x}\in\Omega_K,
\end{align}
where $\beta_{\pi}$ is the inverse temperature and $Z_K$ is the constrained partition function,
\begin{align}
Z_K
=
\sum_{\bm{x}\in\Omega_K}
\exp\left[-\beta_{\pi}E(\bm{x})\right].
\end{align}

We use the Metropolis--Hastings (MH) algorithm to construct the MCMC updates.
Given the current state $\bm{x}$, a candidate state $\bm{y}$ is generated from a proposal distribution $q(\bm{y}\mid\bm{x})$ and accepted with probability
\begin{align}
\alpha(\bm{x},\bm{y})
=
\min\left\{
1,
\frac{
\pi_K(\bm{y})q(\bm{x}\mid\bm{y})
}{
\pi_K(\bm{x})q(\bm{y}\mid\bm{x})
}
\right\}.
\end{align}
If the candidate is rejected, the current state is retained.
By repeating this procedure, the resulting Markov chain converges to the target distribution under appropriate conditions.

As classical baselines, we use global and local Kawasaki dynamics, both of which preserve the Hamming weight by exchanging a pair of bits with different values.
Global Kawasaki dynamics selects a $1$ bit and a $0$ bit from the entire system, whereas local Kawasaki dynamics restricts the exchange to adjacent vertices in the interaction graph.
Both proposal kernels are symmetric, so the MH acceptance probability reduces to
\begin{align}
\alpha(\bm{x},\bm{y})
=
\min\left\{
1,
\exp\left[
-\beta_{\pi}
\left(
E(\bm{y})-E(\bm{x})
\right)
\right]
\right\}.
\end{align}
Because every Kawasaki update preserves the Hamming weight, the Markov chain remains within $\Omega_K$ throughout the sampling process.

\subsection{Quantum Approximate Optimization Algorithm (QAOA)}

The quantum approximate optimization algorithm (QAOA) is a variational quantum algorithm for generating low-cost solutions to combinatorial optimization problems.
For binary variables $\bm{x}\in\{0,1\}^{N}$, a quadratic cost function can be mapped to an Ising Hamiltonian of the form
\begin{align}
H_{\rm C}
=
\sum_{i<j}J_{ij}Z_iZ_j
+
\sum_i h_i Z_i
+
c,
\end{align}
such that
\begin{align}
H_{\rm C}\ket{\bm{x}}
=
E(\bm{x})\ket{\bm{x}}.
\end{align}
For the Boltzmann sampling problems considered in this work, the cost function corresponds to the energy function of the target system.
Here, $J_{ij}$, $h_i$, and $c$ are determined from the QUBO coefficients through the standard binary-to-Ising transformation.

At circuit depth $p$, the QAOA state is
\begin{align}
\ket{\psi(\bm{\gamma},\bm{\beta})}
=
\left(
\prod_{\ell=1}^{p}
e^{-i\beta_\ell H_{\rm M}}
e^{-i\gamma_\ell H_{\rm C}}
\right)
\ket{\psi_0},
\end{align}
where $H_{\rm M}$ is the mixer Hamiltonian and
$\bm{\gamma}=(\gamma_1,\ldots,\gamma_p)$ and
$\bm{\beta}=(\beta_1,\ldots,\beta_p)$
are variational parameters.
The parameters are optimized to minimize the expectation value
\begin{align}
\mathcal{L}(\bm{\gamma},\bm{\beta})
=
\langle
\psi(\bm{\gamma},\bm{\beta})
|
H_{\rm C}
|
\psi(\bm{\gamma},\bm{\beta})
\rangle,
\end{align}
which is estimated from quantum-circuit measurements.
For the fixed Hamming-weight constraint considered in this work, we use the XY mixer \cite{hadfield2019quantum,wang2020xy},
\begin{align}
H_{\rm M}
=
\frac{1}{2}
\sum_{(i,j)\in\mathcal{E}_{\rm M}}
\left(
X_iX_j
+
Y_iY_j
\right).
\end{align}
The XY mixer preserves the total Hamming weight.
Therefore, an initial state in the Hamming-weight-$K$ subspace remains within the feasible subspace throughout the QAOA circuit.

\section{Hardware-Efficient Neural-Surrogated QeMCMC}
\label{subsec:optimization-efficient-qemcmc}

We develop a hardware-efficient implementation of neural-surrogated QeMCMC for large-scale systems.
Our objective is to sample from the constrained Boltzmann distribution $\pi_K(\bm{x})$ defined in Eq.~\eqref{eq:constrained-boltzmann}, with the Markov chain restricted to the feasible set $\Omega_K$.
In the divide-and-conquer approach, a large system is divided into smaller subproblems, and the QAOA sample distributions for each subproblem are learned by separate MADE models and used as MCMC proposal distributions \cite{kawamata2026divide}.
However, independently optimizing QAOA parameters for every subproblem requires repeated expectation-value evaluations and therefore a large number of quantum-circuit executions.
To reduce this cost, we employ linear-ramp QAOA and construct subproblems so that QAOA parameters can be efficiently transferred between them.
As a result, separate QAOA optimization for each subproblem can be avoided, substantially reducing the number of quantum-circuit executions required on quantum hardware.

\begin{figure*}[t]
    \centering
    \includegraphics[width=0.95\textwidth]{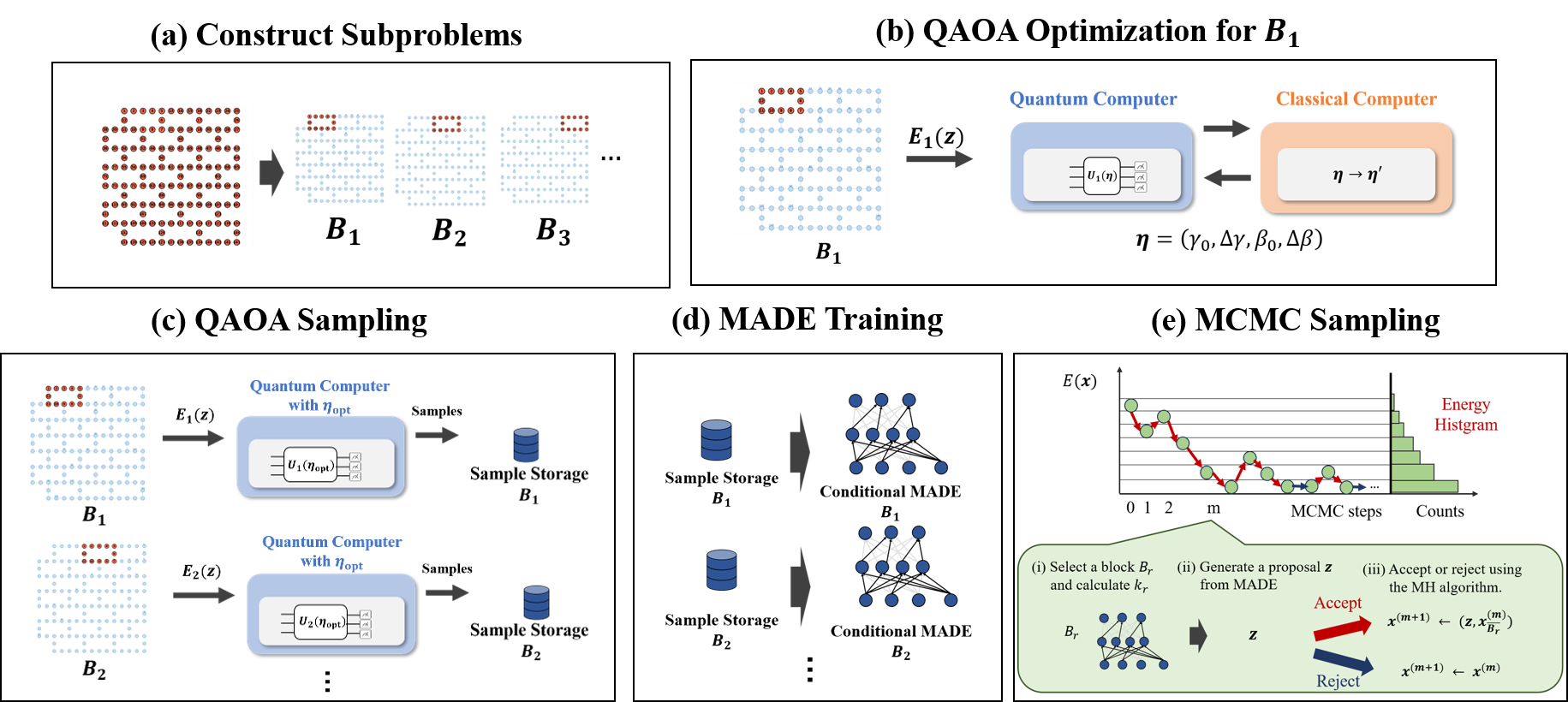}
    \caption{
    Overview of the proposed implementation of neural-surrogated QeMCMC.
    (a) The full system is divided into overlapping subproblems.
    (b) The four parameters of linear-ramp QAOA are optimized for a representative subproblem $B_1$ using a quantum computer and a classical optimizer.
    (c) The optimized parameters are reused to generate QAOA samples for all subproblems without further optimization.
    (d) The samples from each subproblem are used to train a conditional MADE.
    (e) At each MCMC step, a block $B_r$ is selected, and a new configuration is proposed by the corresponding MADE conditioned on the block Hamming weight $k_r$.
    Proposals that violate the Hamming weight constraint are rejected, while valid proposals are accepted or rejected using the Metropolis--Hastings algorithm.}
    \label{fig:algorithm-flow}
\end{figure*}

Figure~\ref{fig:algorithm-flow} shows an overview of the proposed implementation.
For the interaction graph $G=(V,\mathcal{E})$ introduced in Sec.~\ref{sec:preliminaries}, let $\mathcal{B}=\{B_1,\ldots,B_M\}$ denote a collection of vertex blocks satisfying $B_r\subseteq V$ and $\bigcup_{r=1}^{M}B_r=V$.
Each block defines a subproblem.
We allow neighboring blocks to overlap so that successive block updates can propagate changes across block boundaries and throughout the full system.
At the same time, to facilitate QAOA parameter transfer, the subproblems are constructed to have similar local graph structures \cite{galda2021transferability,galda2023similarity}.
For general graph structures, constructing such subproblems may require problem-dependent heuristics that balance these two requirements.
In this work, however, we focus on hardware-native subproblems constructed according to the topology of the IBM quantum processor.
This allows us to explicitly construct subproblems with similar topologies while maintaining the required overlaps, without introducing a dedicated partitioning heuristic.

For each block $B_r$, we define the local cost function using the interactions contained within the block,
\begin{align}
E_r(\bm{z})
&=
\sum_{(i,j)\in\mathcal{E}_r}
Q_{ij}z_i z_j
+
\sum_{i\in B_r}q_i z_i,
\\
\mathcal{E}_r
&=
\left\{
(i,j)\in\mathcal{E}
\,\middle|\,
i,j\in B_r
\right\},
\end{align}
where $\bm{z}\in\{0,1\}^{|B_r|}$ is a configuration of the variables in $B_r$, and $Q_{ij}$ and $q_i$ are the QUBO coefficients defined in Sec.~\ref{sec:preliminaries}.
Interactions crossing the block boundary are omitted from this local cost function.
They are nevertheless included in the full-system energy used for the MCMC acceptance test below.

The corresponding cost Hamiltonian $H_{\rm C}^{(r)}$ is obtained through the same binary-to-Ising transformation introduced in Sec.~\ref{sec:preliminaries} and satisfies
\begin{align}
H_{\rm C}^{(r)}
\ket{\bm{z}}
&=
E_r(\bm{z})
\ket{\bm{z}}.
\end{align}
The block mixer $H_{\rm M}^{(r)}$ has the XY form introduced in Sec.~\ref{sec:preliminaries}, with its mixer graph defined on the vertices of $B_r$.

We first optimize the QAOA parameters only for the first subproblem $B_1$.
To reduce the dimension of the optimization problem, we employ linear-ramp QAOA \cite{montanez2025linear}.
For a fixed circuit depth $p>1$, the angles are parameterized as
\begin{align}
\gamma_\ell
&=
\gamma_0+\Delta\gamma\,u_\ell,
\\
\beta_\ell
&=
\beta_0+\Delta\beta\,(1-u_\ell),
\\
u_\ell
&=
\frac{\ell-1}{p-1},
\qquad
\ell=1,\ldots,p.
\end{align}
The four independent parameters are collected into
$\bm{\eta}=(\gamma_0,\Delta\gamma,\beta_0,\Delta\beta)$,
which determines the angle vectors
$\bm{\gamma}(\bm{\eta})$ and $\bm{\beta}(\bm{\eta})$.

The QAOA state for $B_1$ is
\begin{align}
\ket{\psi_1(\bm{\eta})}
&=
\left(
\prod_{\ell=1}^{p}
e^{-i\beta_\ell H_{\rm M}^{(1)}}
e^{-i\gamma_\ell H_{\rm C}^{(1)}}
\right)
\ket{+}^{\otimes \lvert B_1\rvert},
\end{align}
where the angles are determined by $\bm{\eta}$ through the linear schedules above.
The product is ordered so that layer $\ell=1$ acts first.
The parameters are optimized by minimizing the expected local energy,
\begin{align}
\bm{\eta}_{\rm opt}
&=
\underset{\bm{\eta}}{\arg\min}
\;
\bra{\psi_1(\bm{\eta})}
H_{\rm C}^{(1)}
\ket{\psi_1(\bm{\eta})}.
\end{align}
Thus, the optimized angle vectors
$\bm{\gamma}_{\rm opt}=\bm{\gamma}(\bm{\eta}_{\rm opt})$
and
$\bm{\beta}_{\rm opt}=\bm{\beta}(\bm{\eta}_{\rm opt})$
are obtained by optimizing only four parameters on the first subproblem.

Motivated by the transferability of QAOA parameters between problem instances with similar graph structures and objective landscapes \cite{galda2021transferability,galda2023similarity}, we reuse $\bm{\gamma}_{\rm opt}$ and $\bm{\beta}_{\rm opt}$ for all subproblems $B_r$, $r=1,\ldots,M$.
For each block, the corresponding QAOA state is
\begin{align}
\ket{\psi_r}
&=
\left(
\prod_{\ell=1}^{p}
e^{-i\beta_{{\rm opt},\ell} H_{\rm M}^{(r)}}
e^{-i\gamma_{{\rm opt},\ell} H_{\rm C}^{(r)}}
\right)
\ket{+}^{\otimes \lvert B_r\rvert}.
\end{align}
We measure each state $\ket{\psi_r}$ in the computational basis and use the resulting bit strings as quantum samples for the corresponding block.

The initial state $\ket{+}^{\otimes \lvert B_r\rvert}$ contains components in all block Hamming-weight sectors.
The XY mixer preserves each Hamming-weight sector during QAOA evolution, but the resulting samples can still have different block Hamming weights.
This allows the samples to be used for learning distributions conditioned on different block Hamming weights, as required when the configuration outside the block changes during MCMC.

These samples are then distilled into a conditional MADE.
Each sample $\bm{z}$ is labeled by its measured Hamming weight
$k=\sum_{i\in B_r}z_i$.
For the $a$th autoregressive ordering
$(i_1^{(a)},\ldots,i_{|B_r|}^{(a)})$
of the block variables, the learned conditional likelihood is represented as
\begin{align}
q_{\theta_r,a}^{(r)}(\bm{z}\mid k)
&=
\prod_{t=1}^{\lvert B_r\rvert}
q_{\theta_r,a}^{(r)}
\left(
z_{i_t^{(a)}}
\mid
\bm{z}_{<t}^{(a)},k
\right),
\end{align}
where $\theta_r$ denotes the model parameters and $\bm{z}_{<t}^{(a)}$ denotes the values of the variables preceding $i_t^{(a)}$ in that ordering.
We use the uniform mixture over autoregressive orderings, $q_{\theta_r}^{(r)}(\bm{z}\mid k)$, defined in Eq.~\eqref{eq:made-order-mixture}, as the proposal distribution.

During MCMC, one block $B_r$ is selected at each step.
We assume that the block-selection probabilities are independent of the current state.
Let $\bar{B}_r=V\setminus B_r$ denote the complement of the selected block.
For the current state $\bm{x}\in\Omega_K$, the Hamming weight inside $B_r$ is
\begin{align}
k_r
&=
\sum_{i\in B_r}x_i
=
K-\sum_{i\in\bar{B}_r}x_i.
\end{align}
A proposed block configuration $\bm{z}$ is generated from
$q_{\theta_r}^{(r)}(\cdot\mid k_r)$.
The conditional MADE does not enforce the Hamming-weight constraint by construction.
If the generated configuration satisfies
\begin{align}
\sum_{i\in B_r}z_i
&=
k_r,
\end{align}
the candidate state $\bm{y}$ is defined by
\begin{align}
y_i
&=
\begin{cases}
z_i, & i\in B_r,\\
x_i, & i\in\bar{B}_r.
\end{cases}
\end{align}
Otherwise, the proposal is rejected and the Markov chain remains at the current state.

For a valid candidate state $\bm{y}\in\Omega_K$, the Metropolis--Hastings acceptance probability conditional on selecting block $B_r$ is
\begin{align}
\alpha_r(\bm{x},\bm{y})
&=
\min
\left\{
1,
\frac{
\pi_K(\bm{y})
q_{\theta_r}^{(r)}
\left(\bm{x}_{B_r}\mid k_r\right)
}{
\pi_K(\bm{x})
q_{\theta_r}^{(r)}
\left(\bm{y}_{B_r}\mid k_r\right)
}
\right\}
\\
&=
\min
\left\{
1,
e^{-\beta_{\pi}[E(\bm{y})-E(\bm{x})]}
\frac{
q_{\theta_r}^{(r)}
\left(\bm{x}_{B_r}\mid k_r\right)
}{
q_{\theta_r}^{(r)}
\left(\bm{y}_{B_r}\mid k_r\right)
}
\right\}.
\end{align}
Here, $E(\bm{x})$ is the full-system QUBO energy defined in Sec.~\ref{sec:preliminaries}, rather than the local cost $E_r$ used to generate the quantum samples.
The energy difference therefore includes interactions between the updated block and the unchanged variables outside it.

Because the variables outside $B_r$ remain unchanged, the conditioning value $k_r$ is identical for the forward and reverse proposals.
The block-wise MH update therefore preserves $\pi_K$, and selecting blocks with state-independent probabilities preserves the same target distribution.
Consequently, the MH correction preserves the target distribution even when the learned MADE does not exactly reproduce the QAOA sample distribution or the local QAOA distribution differs from the target conditional distribution.
Convergence to $\pi_K$ additionally requires the usual irreducibility and aperiodicity conditions on the resulting Markov chain within $\Omega_K$.

This implementation substantially reduces the cost of QAOA parameter optimization required to construct neural-surrogated QeMCMC.
Linear-ramp QAOA uses only four independent variational parameters instead of optimizing the $2p$ angles of standard depth-$p$ QAOA separately.
Furthermore, instead of optimizing QAOA parameters independently for all $M$ subproblems, we optimize them only for one representative subproblem and transfer the optimized parameters to the remaining subproblems, reducing the number of independent optimizations from $M$ to one.
Together, these two reductions significantly decrease the number of repeated expectation-value evaluations and quantum-circuit executions required on quantum hardware, making large-scale divide-and-conquer QeMCMC more practical.
This efficiency, however, relies on the transferred QAOA parameters remaining effective for the other subproblems, which we examine experimentally on real quantum hardware in the following section.

\section{Block-Size Scaling Experiment on Quantum Hardware}
\label{sec:experiment-qaoa-made}

\subsection{Experimental Setup}
\label{subsec:qaoa-made-problem}

In this experiment, we implement the hardware-efficient neural-surrogated QeMCMC introduced in Sec.~\ref{subsec:optimization-efficient-qemcmc} to sample from the constrained Boltzmann distribution $\pi_K(\bm{x})$ defined in Sec.~\ref{sec:preliminaries}, using hardware-generated QAOA samples.
For the quantum hardware experiments, we use the IBM Quantum System Two ibm kobe installed at the RIKEN Center for Computational Science (R-CCS) \cite{ibm_riken_system_two}.
The system contains an IBM Quantum Heron r2 processor with 156 superconducting qubits connected through a sparse heavy-hex graph \cite{wang2026markovian,xu2026filter,shinjo2026synchronization}.

\begin{figure}[t]
    \centering
    \includegraphics[width=0.6\columnwidth]{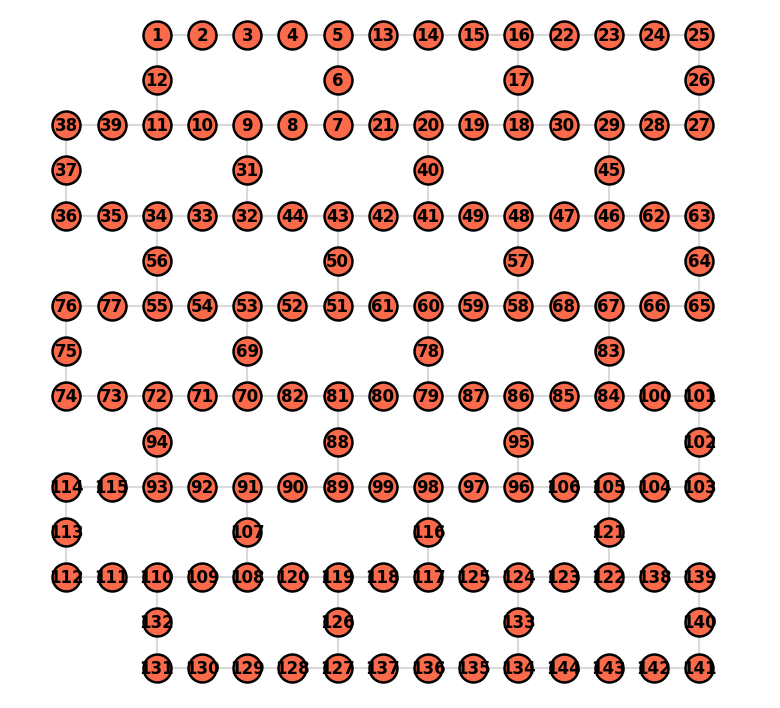}
    \caption{
    Problem topology used in the hardware experiment.
    The interaction graph is constructed to be compatible with the
    hardware connectivity of the IBM Quantum Heron r2 processor.
    }
    \label{fig:qaoa-made-problem-topology}
\end{figure}

We use the QUBO energy function $E(\bm{x})$ introduced in Sec.~\ref{sec:preliminaries} for a system of $N=144$ binary variables.
The interaction graph $G=(V,\mathcal{E})$ is constructed to match the hardware connectivity of the IBM Quantum Heron r2 processor, as shown in Fig.~\ref{fig:qaoa-made-problem-topology}.
Matching the QUBO interaction graph to the hardware connectivity suppresses routing overhead during circuit execution \cite{pelofske2023quantum,montanez2025evaluating,bucher2026constrained}.
For each edge $(i,j)\in\mathcal{E}$, the quadratic coefficient $Q_{ij}$ is independently drawn from a standard normal distribution, $Q_{ij}\sim\mathcal{N}(0,1)$, using a fixed random seed for reproducibility.
The linear coefficients are set to $q_i=0$.
We focus on the half-filled case and set the total Hamming weight to $K=N/2$.

To investigate how the MCMC performance scales with the number of variables updated at once, we vary block sizes while keeping the full system size fixed at $N=144$.
Let $b=|B_r|$ denote the block size, i.e., the number of binary variables in each block $B_r$.
We consider
\begin{align}
b
\in
\{2,12,30,49,82,144\}.
\label{eq:block-size-sweep}
\end{align}
The subproblems are constructed to share some vertices across blocks and to have similar graph topologies to facilitate QAOA parameter transfer.
Figure~\ref{fig:algorithm-flow} (1) shows an example of the partitioning for $b=12$.
For each block size, we generate QAOA samples using the parameter-transfer procedure, distill them into conditional MADE surrogates, and evaluate the MCMC performance using the resulting block-wise proposal distributions.
Additional details on MADE training, quantum-sampling budgets, and block definitions are provided in Appendix~\ref{app:experiment_details}.

\begin{figure*}[h]
    \centering
    \includegraphics[width=0.9\linewidth]{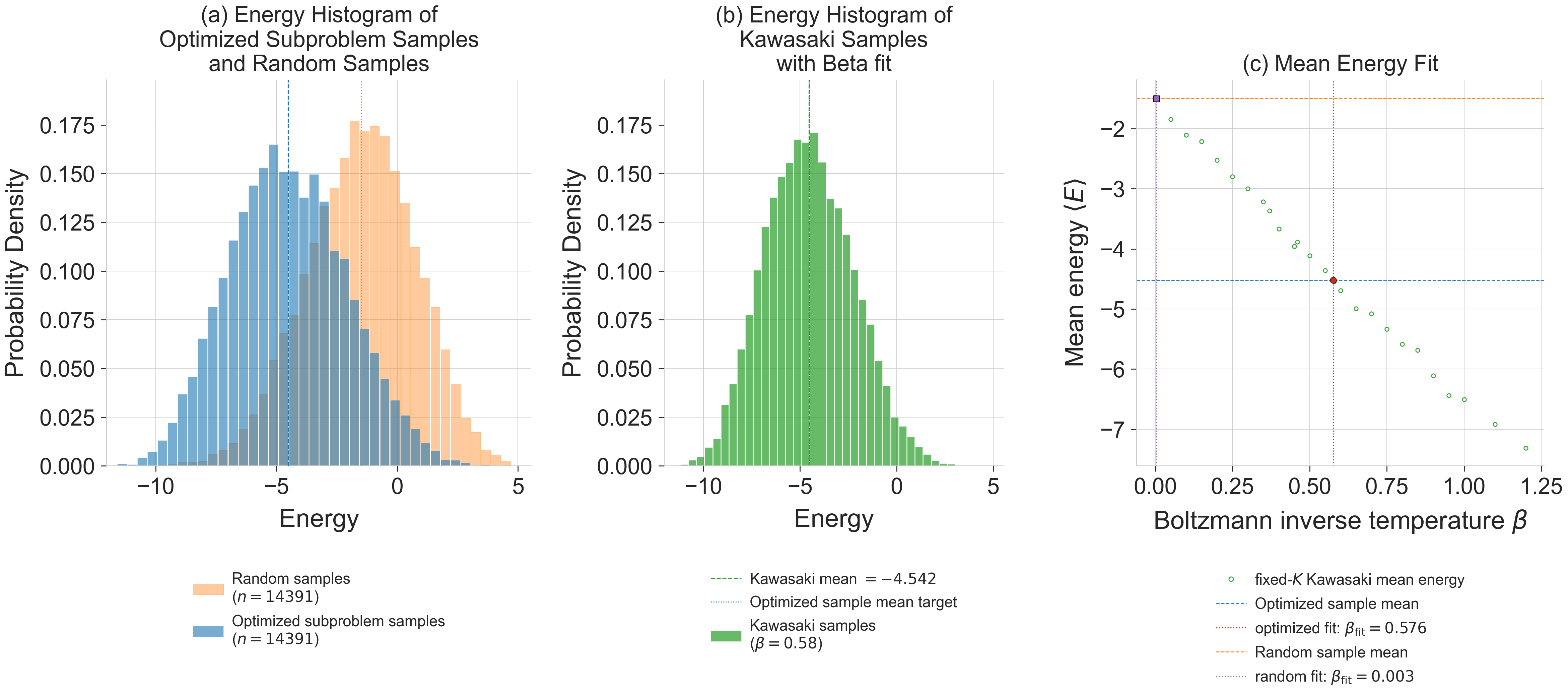}
    \caption{Example of the procedure for estimating the effective inverse temperature $\beta_{\rm fit}$ of QAOA samples. For the selected subproblem, Boltzmann samples are generated using Kawasaki dynamics at several inverse temperatures $\beta$, and their mean energies are compared with the mean energy of the QAOA samples. The fitted inverse temperature $\beta_{\rm fit}$ is defined as the value of $\beta$ whose mean energy is closest to that of the QAOA samples. The same procedure is applied to uniformly random samples for comparison, for which the fitted value is expected to be close to the infinite-temperature limit $\beta=0$.}
    \label{fig:qaoa_kawasaki_beta_fit}
\end{figure*}

\begin{figure*}[h]
    \centering
    \includegraphics[width=0.9\linewidth]{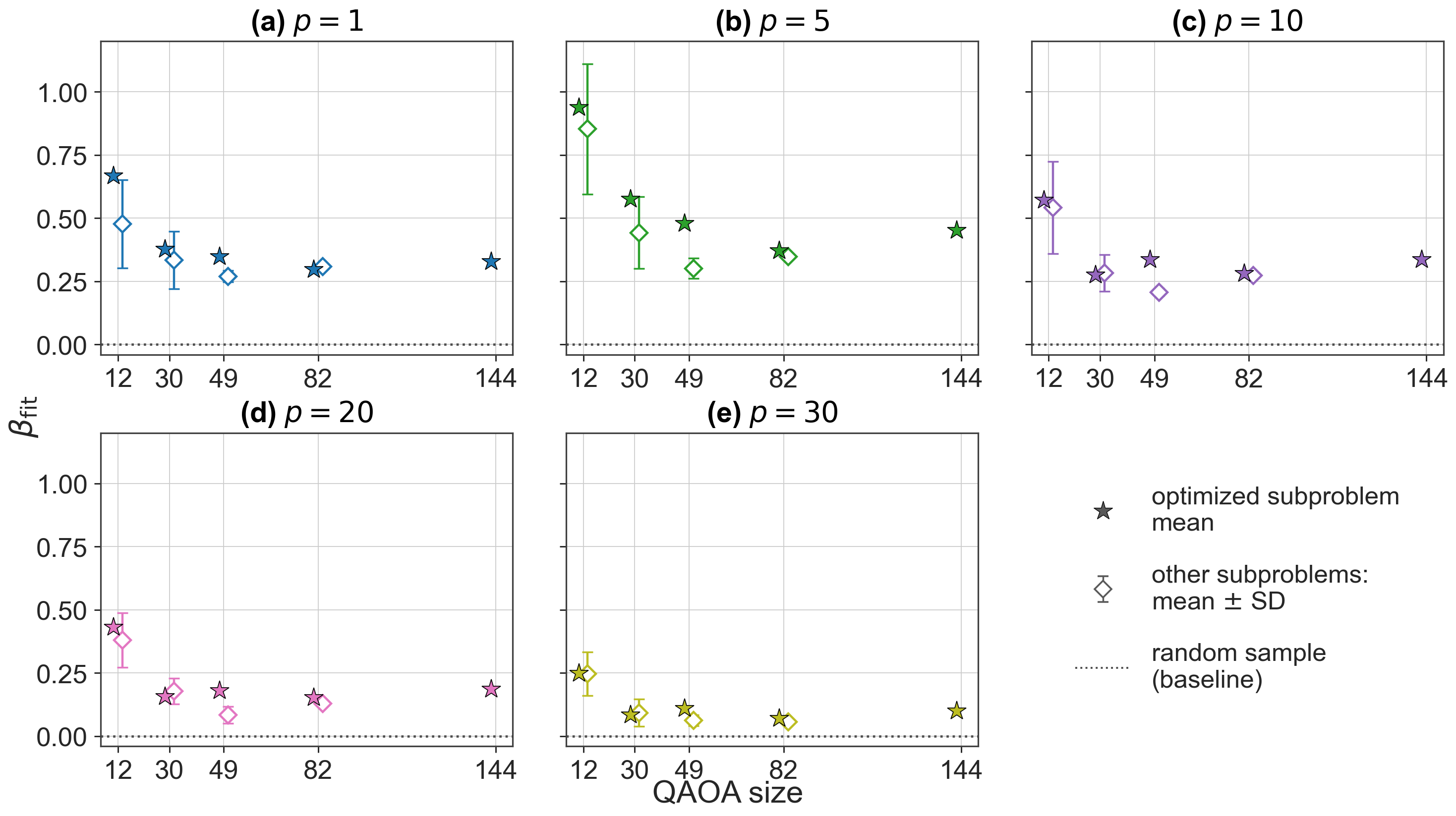}
    \caption{Effective inverse temperature $\beta_{\rm fit}$ of the QAOA sample distributions as a function of the block size $b$ for different QAOA depths $p$. For each block size, the star denotes the result for the representative subproblem on which the QAOA parameters are optimized. The diamonds show the mean $\beta_{\rm fit}$ over the remaining subproblems obtained using the transferred QAOA parameters, and the error bars indicate the corresponding standard deviation across these subproblems. The effective inverse temperature is estimated by comparing the mean energy of the QAOA samples with that of a Boltzmann distribution for the same subproblem and Hamming weight sector. Uniform sampling corresponds to $\beta=0$, whereas a positive $\beta_{\rm fit}$ indicates a lower effective temperature.}
    \label{fig:beta_fit_vs_block_size}
\end{figure*}

\subsection{QAOA Results}
\label{subsec:qaoa-results}

We first examine whether QAOA can still generate samples at low effective temperatures when the optimized parameters are transferred to other subproblems.
To assess this, we compare the effective temperatures of the samples from the representative subproblem and the other subproblems, using uniform sampling as a reference.
Previous studies have shown that QAOA can generate distributions that approximately resemble Boltzmann distributions \cite{diez2023quantum,lotshaw2023approximate}.
We therefore estimate an effective inverse temperature $\beta_{\rm fit}$ from the mean energy of the samples.

Figure~\ref{fig:qaoa_kawasaki_beta_fit} shows a representative procedure for estimating $\beta_{\rm fit}$.
For each target subproblem, we generate Boltzmann samples using Kawasaki dynamics at several inverse temperatures $\beta$ and evaluate their mean energies.
We then define $\beta_{\rm fit}$ as the value of $\beta$ whose mean energy is closest to that of the QAOA samples in the same fixed-Hamming-weight sector.
For comparison, the same fitting procedure is also applied to uniform samples.
As shown in the figure, the fitted value for uniform sampling is close to $\beta=0$, consistent with the fact that a uniform random distribution corresponds to the infinite-temperature limit, $\beta=0$.
Thus, a positive $\beta_{\rm fit}$ indicates a lower effective temperature than uniform sampling, based on the mean energy within the same subproblem and Hamming weight sector.

Figure~\ref{fig:beta_fit_vs_block_size} shows $\beta_{\rm fit}$ for different block sizes $b$ and QAOA depths $p$.
For each QAOA depth $p$, $\beta_{\rm fit}$ is relatively large for small blocks and decreases as the block size increases.
For larger block sizes, however, the dependence on $b$ becomes weaker and $\beta_{\rm fit}$ approaches an approximately constant value.
This behavior is qualitatively consistent with previous hardware-native QAOA studies, which reported that the system-size dependence of shallow-QAOA performance becomes weak for sufficiently large sparse graphs \cite{harrigan2021quantum,pelofske2023quantum}.
The effective inverse temperatures show similar trends for the representative subproblem and the other subproblems using transferred parameters, across different subproblem sizes and QAOA depths.
Although the values vary among subproblems, they generally remain positive.
These results suggest that parameter transfer preserves the ability of QAOA to generate samples at lower effective temperatures than uniform sampling, without separate parameter optimization
for each subproblem.
In most of the tested settings, the estimated effective inverse temperatures are below $0.5$.
Based on these results, we evaluate MCMC at higher target inverse temperatures $\beta_{\pi}=0.5,\,0.7,\,0.9,$ and $1.2$ in the next subsection.

\subsection{MCMC Results}
\label{subsec:qaoa-made-results}

\begin{figure}[!htbp]
    \centering
    \includegraphics[width=0.9\linewidth]{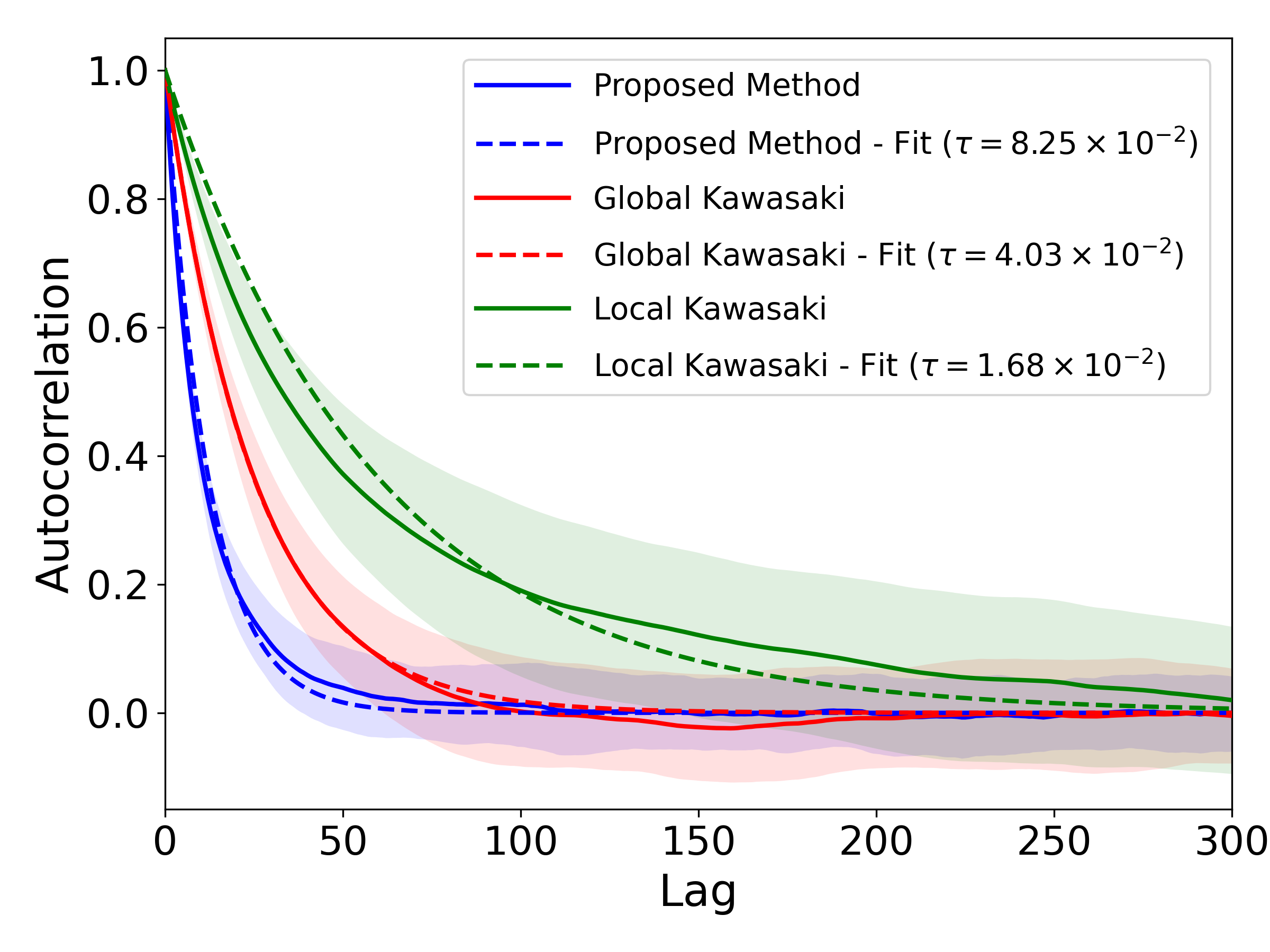}
    \caption{
    Representative extraction of the autocorrelation decay rate $\tau$
    for $N=144$, $K=72$, subproblem size $b=12$,
    QAOA depth $p=20$, and target inverse temperature $\beta_{\pi}=0.5$.
    The horizontal axis is the lag $\ell$ in MCMC steps,
    and the vertical axis is the normalized autocorrelation averaged over replica pairs.
    The Hamming-distance autocorrelation is equivalent to the replica-overlap autocorrelation.
    Solid lines show the measured autocorrelations for the proposed method,
    global Kawasaki dynamics, and local Kawasaki dynamics.
    Dashed lines show fits to $\overline{C}(\ell)=\exp(-\tau\ell)$,
    from which $\tau$ is estimated.
    }
    \label{fig:autocorrelation-tau-fit}
\end{figure}

\begin{figure*}[!htbp]
    \centering
    \includegraphics[width=0.80\textwidth]{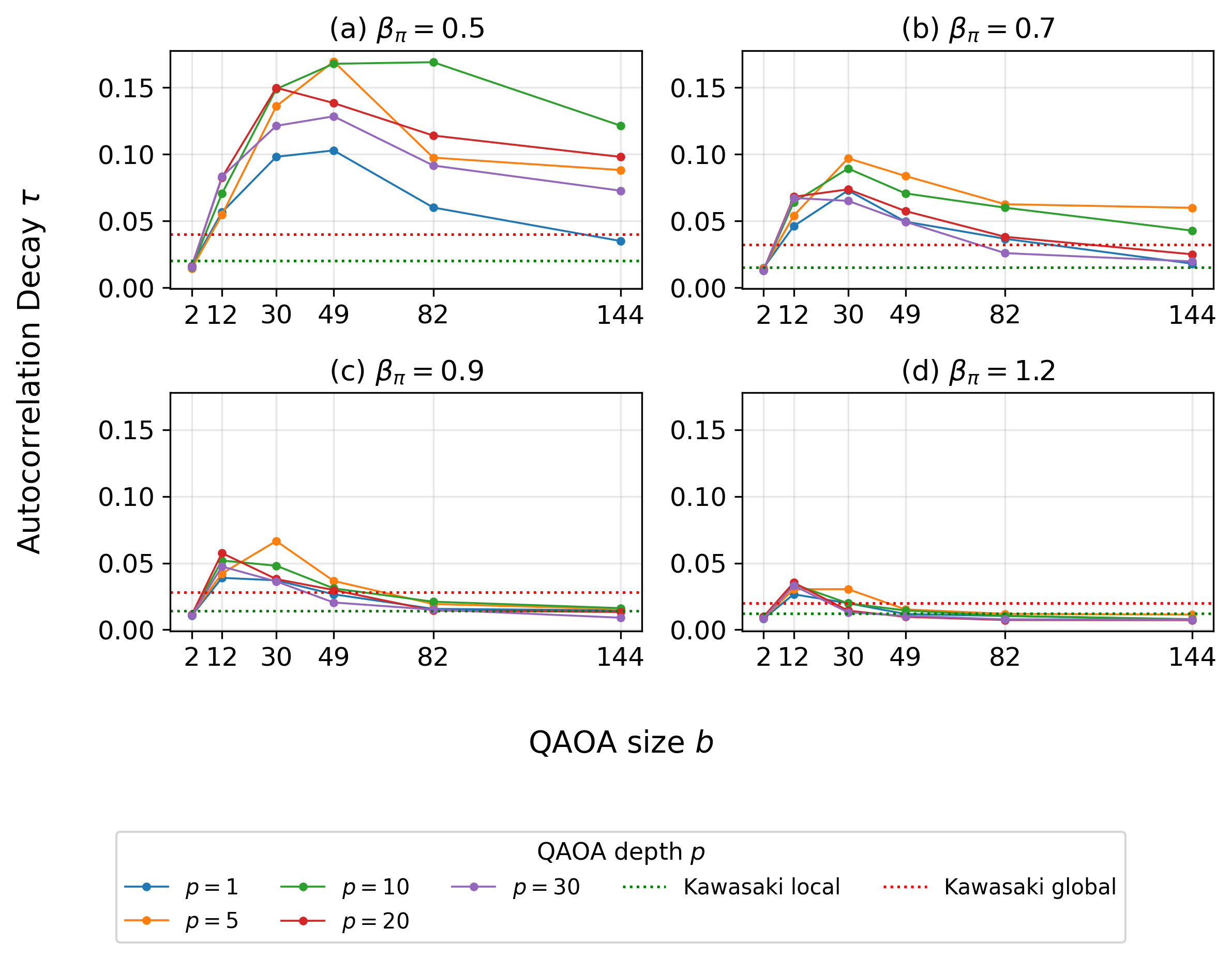}
    \caption{
    Autocorrelation decay rate $\tau$ as a function of the block size $b$ for different QAOA depths $p$ and target inverse temperatures $\beta_{\pi}$.
    The rates are extracted from exponential fits to the replica-pair autocorrelations, as illustrated in Fig.~\ref{fig:autocorrelation-tau-fit}.
    Neural-surrogate proposals trained on hardware-generated QAOA samples are compared with global and local Kawasaki dynamics.
    A larger $\tau$ indicates faster autocorrelation decay per MCMC step.
    }
    \label{fig:autocorrelation-decay}
\end{figure*}

\begin{figure}[!htbp]
    \centering
    \includegraphics[width=0.9\linewidth]{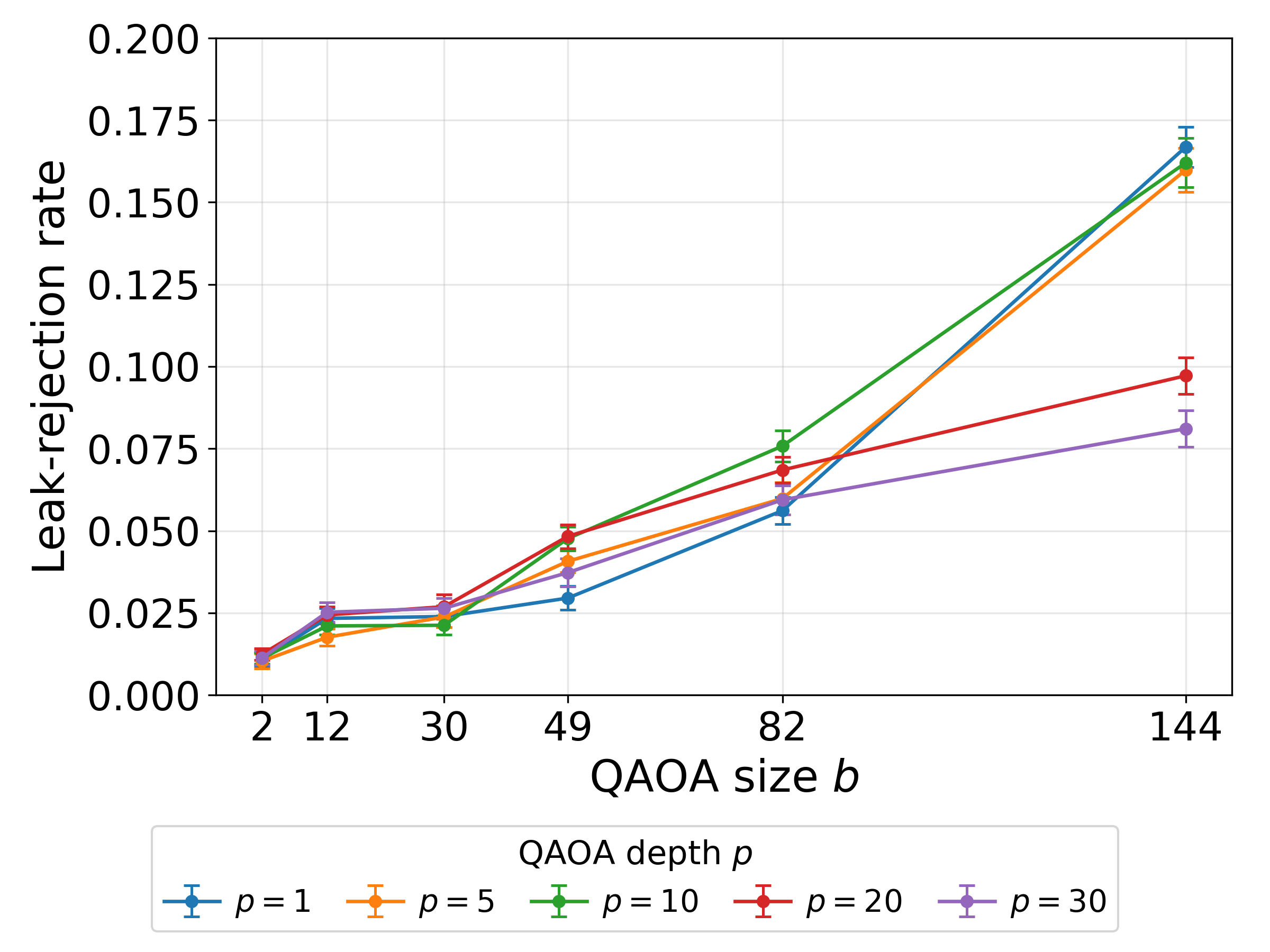}
    \caption{
    Constraint-leakage rates of MADE-generated proposals as a function of the block size $b$, pooled over target inverse temperatures
    $\beta_{\pi}=0.5,\,0.7,\,0.9,\,1.2$.
    Leakage is the fraction of proposals that do not have the required Hamming weight $k_r$, i.e., $\sum_{i\in B_r}z_i\neq k_r$.
    Such proposals are rejected before the Metropolis--Hastings acceptance test.
    Larger blocks exhibit higher leakage rates under the model and training settings used here.}
    \label{fig:mcmc-leakage}
\end{figure}

We evaluate the MCMC mixing performance by varying the block size $b$, QAOA depth $p$, and target inverse temperature $\beta_{\pi}$.
For each setting, we compare the proposed method with global and local Kawasaki dynamics.
Mixing is quantified using the autocorrelation of the normalized Hamming distance between independent replicas.
For each method and target inverse temperature, we run $R=16$ independent replicas for $T=3{,}000$ MCMC steps.
For replicas $\mu$ and $\nu$, this distance at MCMC step $t$ is
\begin{align}
d_{\mu\nu}(t)
&=
\frac{1}{N}
\sum_{i=1}^{N}
\left|
x_i^{(\mu)}(t)-x_i^{(\nu)}(t)
\right|.
\label{eq:mcmc-hamming-distance}
\end{align}
The corresponding replica overlap is
$q_{\mu\nu}(t)=1-2d_{\mu\nu}(t)$.
The two observables therefore have identical normalized autocorrelation functions, making this evaluation equivalent to the replica-overlap analysis used in Ref.~\cite{kawamata2026divide}.
For each replica pair, we calculate
\begin{align}
C_{\mu\nu}(\ell)
&=
\frac{
\sum_{t=0}^{T-\ell-1}
\delta d_{\mu\nu}(t)\,
\delta d_{\mu\nu}(t+\ell)
}{
\sum_{t=0}^{T-1}
[\delta d_{\mu\nu}(t)]^2
},
\label{eq:mcmc-hamming-autocorrelation}
\end{align}
where $\ell$ is the lag in MCMC steps,
$T$ is the number of recorded steps, and
$\delta d_{\mu\nu}(t)=d_{\mu\nu}(t)-\bar{d}_{\mu\nu}$,
with $\bar{d}_{\mu\nu}$ denoting the time-averaged distance.
We average $C_{\mu\nu}(\ell)$ over all replica pairs and fit the resulting mean autocorrelation $\overline{C}(\ell)$ to $\exp(-\tau\ell)$.
Figure~\ref{fig:autocorrelation-tau-fit} shows a representative example of the measured autocorrelation as a function of lag and the corresponding exponential fits.
We use the decay rate $\tau$ as the mixing metric: a larger $\tau$ indicates faster autocorrelation decay per MCMC step.

Figure~\ref{fig:autocorrelation-decay} presents the main MCMC result.
For a range of block sizes, QAOA depths, and target inverse temperatures, the neural-surrogate proposals learned from hardware-generated QAOA samples yield larger $\tau$ than both global and local Kawasaki dynamics.
These results demonstrate that the proposed framework can achieve larger autocorrelation decay rates per MCMC step than those of these classical baselines using samples generated on current quantum hardware.
The improvement is obtained despite hardware noise and the use of transferred QAOA parameters, supporting the practical potential of the hardware-efficient implementation.

The temperature dependence is broadly consistent with the effective inverse temperatures $\beta_{\rm fit}$ of the QAOA samples shown in Fig.~\ref{fig:beta_fit_vs_block_size}.
The target inverse temperatures $\beta_{\pi}$ considered here are generally larger than $\beta_{\rm fit}$, corresponding to targets colder than the effective temperatures inferred for the QAOA samples.
As the target temperature is lowered further, this effective temperature mismatch increases, and the mixing performance generally deteriorates.
This trend is consistent with the proposal distributions becoming less well matched to targets that place more probability on states with lower energies.

An important difference from the previous classical simulations \cite{kawamata2026divide} appears in the dependence on subproblem size.
Increasing $b$ does not always increase the autocorrelation decay rate, even though a larger subproblem allows more variables to be updated in a single proposal.
In contrast, the previous simulations always showed improved mixing as the subproblem size increased over the smaller sizes examined there.
The hardware experiments therefore reveal a block-size-dependent limitation that was not apparent in those small-scale simulations.

Figure~\ref{fig:mcmc-leakage} shows the fraction of MADE-generated proposals that violate the requested block Hamming weight.
To investigate possible reasons for the slower autocorrelation decay at larger subproblem sizes, we examine whether the MADE proposals satisfy the required Hamming weight.
Specifically, a block configuration $\bm{z}$ generated from $q_{\theta_r}^{(r)}(\cdot\mid k_r)$ does not necessarily satisfy $\sum_{i\in B_r}z_i=k_r$.
The fraction of such invalid proposals increases with $b$, and each invalid proposal is rejected without updating the Markov chain.
Because each quantum sample used for training is labeled by its measured Hamming weight, the conditional training data satisfy the specified constraint.
The increasing leakage therefore suggests that reproducing the Hamming-weight-conditioned distributions becomes more difficult for the MADE models at larger block sizes under the present training settings.

This increase in leakage occurs despite a substantial increase in the number of quantum samples.
As summarized in Table~\ref{tab:appA-sampling-made}, we collect
$100{,}000$, $300{,}000$, $500{,}000$, and $1{,}000{,}000$
hardware samples per block for
$b=30$, $49$, $82$, and $144$, respectively.
Thus, the difficulty in reproducing the requested Hamming-weight constraint persists even with large quantum-sampling budgets.

These sample counts, however, are totals across different Hamming-weight sectors and do not by themselves guarantee sufficient training data for every conditioning value.
Although leakage directly causes rejected MCMC steps, it does not by itself establish how much of the observed deterioration in mixing is attributable to surrogate-learning errors.
In particular, the present results do not distinguish insufficient sample coverage from limitations in model capacity or training optimization.
A more detailed investigation is needed to separate these effects and assess their contributions relative to other sources of mismatch between the proposal and target distributions.

Overall, the hardware experiments demonstrate larger autocorrelation decay rates per MCMC step than those of the classical Kawasaki baselines for appropriate block sizes, while showing that larger quantum subproblems do not automatically translate into faster MCMC mixing.
The increase in constraint leakage suggests that neural-surrogate learning is one factor that warrants further investigation when scaling the method.
Supplementary acceptance-rate results are provided in Fig.~\ref{fig:mcmc-acceptance} in Appendix~\ref{app:experiment_details}.

\section{Large-Scale Demonstration on Quantum Hardware}
\label{sec:large-scale-demonstration}

\subsection{Experimental Setup}
\label{subsec:large-scale-setting}

We consider constrained Boltzmann sampling from $\pi_K(\bm{x})$ in Eq.~\eqref{eq:constrained-boltzmann}, with $N=1064$, $K=N/2=532$, and $\beta_{\pi}=0.5$.
The interaction graph is constructed by arranging the $N=144$ brick-pattern graph in Fig.~\ref{fig:qaoa-made-problem-topology} in a $3\times3$ array and identifying vertices and edges shared by neighboring patches.
This produces a connected graph with 1064 vertices and 1252 edges.
The construction extends the same hardware-compatible connectivity pattern used in the preceding experiment.
The quadratic coefficients are drawn from a standard normal distribution, $Q_{ij}\sim\mathcal{N}(0,1)$, using a fixed random seed, and the linear coefficients are set to $q_i=0$.

We choose the block size $b=49$ and QAOA depth $p=10$.
Repeating the four-block decomposition of the $N=144$ graph within each patch yields $M=36$ overlapping blocks.
Each block contains 49 vertices and 54 internal edges and retains a topology compatible with the quantum processor connectivity.
Thus, increasing the full problem size does not require increasing the number of qubits used in an individual QAOA circuit.

Quantum samples are generated using the IBM Quantum Heron r2 processor in ibm kobe, following the same hardware workflow as in Sec.~\ref{sec:experiment-qaoa-made}.
The linear-ramp parameters optimized for the representative $b=49$, $p=10$ subproblem of the $N=144$ experiment are transferred to all 36 blocks.
We collect $300{,}000$ hardware samples per block, corresponding to $10.8$ million samples in total.
The per-block sampling budget, conditional MADE architecture, training settings, and ensemble of 16 autoregressive orderings are the same as those used for $b=49$ in the preceding experiment.

We evaluate mixing using the normalized Hamming-distance autocorrelation defined in Sec.~\ref{subsec:qaoa-made-results}.
For each method, we run $R=16$ independent replicas for $T=27{,}000$ MCMC steps and record the state after every step.
For the analysis below, these replicas are grouped into eight disjoint replica pairs.
We average their autocorrelations and fit the resulting mean $\overline{C}(\ell)$ to $\exp(-\tau\ell)$.
As before, a larger $\tau$ indicates faster autocorrelation decay per MCMC step.

\subsection{Results}
\label{subsec:large-scale-result}

\begin{figure}[t]
    \centering
    \includegraphics[width=0.9\linewidth]{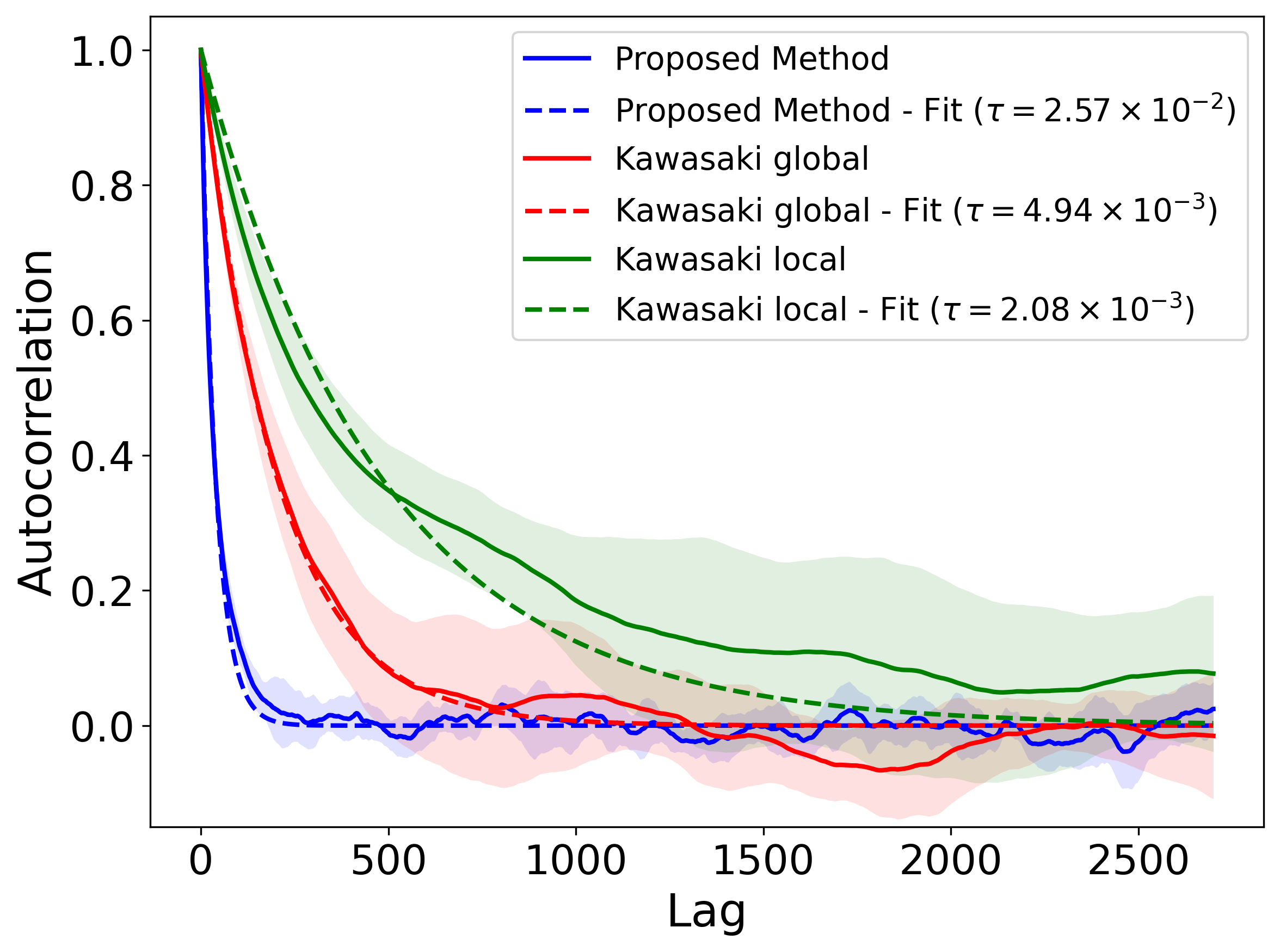}
    \caption{
    Hamming-distance autocorrelation for the $N=1064$ system at $\beta_{\pi}=0.5$, with block size $b=49$ and QAOA depth $p=10$.
    Solid lines show the mean autocorrelations over eight disjoint replica pairs, and shaded regions indicate one standard deviation across these pairs.
    Dashed lines show exponential fits $\overline{C}(\ell)=\exp(-\tau\ell)$.
    The proposed method achieves larger decay rates than both global and local Kawasaki dynamics.
    }
    \label{fig:n1064-mcmc}
\end{figure}

We first characterize the hardware-generated QAOA samples using the mean-energy-matching procedure described in Sec.~\ref{subsec:qaoa-results}.
For each block, the effective inverse temperature is estimated in a fixed-Hamming-weight sector of $k=24$.
Averaging over all 36 blocks yields $\overline{\beta}_{\rm fit}=0.300\pm0.072$, where the uncertainty denotes the sample standard deviation across blocks rather than the standard error of the mean.

Figure~\ref{fig:n1064-mcmc} shows the autocorrelation functions obtained using the learned proposals and the two Kawasaki baselines.
The fitted autocorrelation decay rate per MCMC step is approximately $5.2$ times that of global Kawasaki dynamics and $12.3$ times that of local Kawasaki dynamics.

These improvements are comparable to those obtained for $N=144$ at the same block size, circuit depth, and target inverse temperature, where the corresponding ratios are approximately $4.71$ and $9.95$.
A substantial acceleration is therefore retained when the system size is increased to $N=1064$ while keeping the quantum block size fixed.

This observation is qualitatively consistent with the divide-and-conquer simulations of Ref.~\cite{kawamata2026divide}, which found sustained improvements over Kawasaki dynamics as the system size increased at a fixed block size.
For noiseless simulations on 3-regular graphs with $b=16$, that study reported decay-rate improvements averaged over system sizes of approximately $7.6$ and $20.3$ relative to global and local Kawasaki dynamics, respectively.
Although the graph structure, block size, and hardware conditions differ here, our results show that the same qualitative advantage persists for a system exceeding one thousand variables.
They demonstrate that proposals learned from hardware-generated QAOA samples can accelerate constrained Boltzmann sampling on a large system using fixed-size quantum subproblems.

\section{Conclusion}

In this work, we developed a hardware-efficient implementation of neural-surrogated QeMCMC for constrained Boltzmann sampling.
By combining linear-ramp QAOA with parameter transfer between subproblems with similar graph topologies, the implementation requires optimizing only four parameters on a representative subproblem and reuses them for the remaining subproblems.
This avoids independent parameter optimization for every subproblem and substantially reduces the number of quantum-circuit executions required for parameter optimization.

Using an IBM Quantum Heron r2 processor, we first investigated the block-size dependence of MCMC performance for a system of $N=144$ binary variables.
Parameter transfer generally preserved the ability of QAOA to generate samples at lower effective temperatures than uniform sampling.
For suitable subproblem sizes, the learned proposals yielded larger autocorrelation decay rates per MCMC step than those of both global and local Kawasaki dynamics.
By varying the block size up to $b=144$, we observed a deterioration in mixing at larger block sizes that was not apparent in the previous small-scale simulations.
The accompanying increase in constraint leakage from MADE-generated proposals suggests difficulties in learning high-dimensional conditional distributions under the present model and training settings.
However, leakage alone does not establish surrogate learning as the dominant cause of the deterioration in mixing.

Building on these observations, we extended the full system to $N=1064$ while fixing the block size at $b=49$, for which the learned proposals had yielded larger autocorrelation decay rates than those of both Kawasaki baselines.
The QAOA parameters optimized for the representative $b=49$, $p=10$ subproblem in the $N=144$ experiment were transferred to all 36 blocks without further optimization.
At $\beta_{\pi}=0.5$, the fitted autocorrelation decay rate per MCMC step was approximately $5.2$ times that of global Kawasaki dynamics and $12.3$ times that of local Kawasaki dynamics.
These improvements were comparable to those obtained for $N=144$ at the same block size, circuit depth, and target inverse temperature.
Thus, substantial acceleration was retained when the full system was enlarged without increasing the number of variables handled by each quantum circuit or neural surrogate.

Future work should disentangle the effects of training-sample coverage, model capacity, and training optimization on constraint leakage and MCMC performance.
These studies should also distinguish surrogate-learning errors from mismatches between the quantum-generated sample distributions and the target conditional distributions, and guide the development of improved neural-network models and training procedures.
Overall, our results demonstrate that proposals learned from hardware-generated samples of fixed-size quantum subproblems can increase the autocorrelation decay rate per MCMC step for constrained Boltzmann sampling in a system exceeding one thousand variables.

\begin{acknowledgments}
This work was supported by MEXT Quantum Leap Flagship Program (MEXT Q-LEAP) Grant No. JPMXS0120319794, JST COI-NEXT Grant No. JPMJPF2014, and JST CREST Grant No. JPMJCR24I3. This paper is based on results obtained from a project, JPNP20017, commissioned by the New Energy and Industrial Technology Development Organization (NEDO).
\end{acknowledgments}

\appendix

\section{Experimental Details}
\label{app:experiment_details}
\label{app:made-training}
\label{app:made-orderings}
\label{app:subproblem-partitions}

For each block $B_r$, we train a conditional MADE based on the architecture of Ref.~\cite{germain2015made}, using hardware-generated QAOA samples obtained through the procedure described in Sec.~\ref{subsec:optimization-efficient-qemcmc}.
The samples are split into 90\% for training and 10\% for validation, and the model is trained until the loss converges.
To condition the model on the block Hamming weight $k$, we supply a one-hot representation through auxiliary context units connected to the hidden layers: the unit corresponding to $k$ is set to $1$, and all others are set to $0$.
We use 145 context units for $b\geq12$ and three for $b=2$.
Table~\ref{tab:appA-sampling-made} lists the number of quantum samples collected per block, $S_{\rm sample}$, and the number of autoregressive orderings, $A_r$.
The sample counts are totals across block Hamming-weight sectors, and the sampling budget is increased for larger blocks.

We combine multiple autoregressive orderings to reduce dependence on a particular variable ordering \cite{germain2015made}.
For each block $B_r$, the proposal distribution is the uniform mixture
\begin{align}
q_{\theta_r}^{(r)}(\bm{z}\mid k)
&=
\frac{1}{A_r}
\sum_{a=1}^{A_r}
q_{\theta_r,a}^{(r)}(\bm{z}\mid k),
\label{eq:made-order-mixture}
\end{align}
where $q_{\theta_r,a}^{(r)}$ denotes the conditional distribution associated with the $a$th ordering.
A proposal is generated by selecting an ordering uniformly and sampling autoregressively from the corresponding distribution.
The forward and reverse proposal probabilities in the Metropolis--Hastings acceptance test are evaluated using the full mixture likelihood in Eq.~\eqref{eq:made-order-mixture}.

\begin{table}[t]
    \caption{
    QAOA sampling budgets and numbers of autoregressive orderings for each block size $b$.
    $S_{\rm sample}$ is the total number of hardware samples collected per block across Hamming-weight sectors, and $A_r$ is the number of orderings in the proposal ensemble.
    }
    \label{tab:appA-sampling-made}
    \begin{ruledtabular}
    \begin{tabular}{crr}
    $b$ & $S_{\rm sample}$ & $A_r$ \\
    \hline
    2   & $5{,}000$       & 2 \\
    12  & $30{,}000$      & 3 \\
    30  & $100{,}000$     & 3 \\
    49  & $300{,}000$     & 16 \\
    82  & $500{,}000$     & 16 \\
    144 & $1{,}000{,}000$ & 32 \\
    \end{tabular}
    \end{ruledtabular}
\end{table}

Table~\ref{tab:subproblem-partitions} lists the vertex sets defining the blocks $B_r$ for each block size $b$.
The vertex indices refer to the binary variables of the full $N=144$ system, and the family labels A and B distinguish the block collections.
These blocks are used in the parameter-transfer and block-wise MCMC procedures described in Sec.~\ref{subsec:optimization-efficient-qemcmc}.

\begin{table*}[t]
    \caption{
    Vertex sets used for the subproblem partitions of the $N=144$ system.
    Each set specifies a block $B_r$ of size $b$.
    The notation $u{:}v$ denotes $\{u,u+1,\ldots,v\}$, and semicolons separate blocks.
    A and B denote the block families.
    }
    \label{tab:subproblem-partitions}
    \scriptsize
    \squeezetable
    \begin{ruledtabular}
    \begin{tabular}{ccp{0.83\textwidth}}
    $b$ & Family & Vertex sets \\
    \hline

    2 & A &
    $\{1,2\}$; $\{1,12\}$; $\{3,4\}$; $\{5,6\}$; $\{5,13\}$;
    $\{7,8\}$; $\{7,21\}$; $\{9,31\}$; $\{10,11\}$; $\{11,39\}$;
    $\{14,15\}$; $\{16,17\}$; $\{16,22\}$; $\{18,19\}$; $\{18,30\}$;
    $\{20,40\}$; $\{23,24\}$; $\{25,26\}$; $\{27,28\}$; $\{29,45\}$;
    $\{32,33\}$; $\{32,44\}$; $\{34,56\}$; $\{35,36\}$; $\{37,38\}$;
    $\{41,42\}$; $\{41,49\}$; $\{43,50\}$; $\{46,47\}$; $\{46,62\}$;
    $\{48,57\}$; $\{51,52\}$; $\{51,61\}$; $\{53,69\}$; $\{54,55\}$;
    $\{55,77\}$; $\{58,59\}$; $\{58,68\}$; $\{60,78\}$; $\{63,64\}$;
    $\{65,66\}$; $\{67,83\}$; $\{70,71\}$; $\{70,82\}$; $\{72,94\}$;
    $\{73,74\}$; $\{75,76\}$; $\{79,80\}$; $\{79,87\}$; $\{81,88\}$;
    $\{84,85\}$; $\{84,100\}$; $\{86,95\}$; $\{89,90\}$; $\{89,99\}$;
    $\{91,107\}$; $\{92,93\}$; $\{93,115\}$; $\{96,97\}$; $\{96,106\}$;
    $\{98,116\}$; $\{101,102\}$; $\{103,104\}$; $\{105,121\}$;
    $\{108,109\}$; $\{108,120\}$; $\{110,132\}$; $\{111,112\}$;
    $\{113,114\}$; $\{117,118\}$; $\{117,125\}$; $\{119,126\}$;
    $\{122,123\}$; $\{124,133\}$; $\{127,137\}$; $\{128,129\}$;
    $\{130,131\}$; $\{134,144\}$; $\{135,136\}$; $\{138,139\}$;
    $\{140,141\}$; $\{142,143\}$
    \\[2pt]

    2 & B &
    $\{2,3\}$; $\{4,5\}$; $\{6,7\}$; $\{8,9\}$; $\{9,10\}$;
    $\{11,12\}$; $\{13,14\}$; $\{15,16\}$; $\{17,18\}$; $\{19,20\}$;
    $\{20,21\}$; $\{22,23\}$; $\{24,25\}$; $\{26,27\}$; $\{28,29\}$;
    $\{29,30\}$; $\{31,32\}$; $\{33,34\}$; $\{34,35\}$; $\{36,37\}$;
    $\{38,39\}$; $\{40,41\}$; $\{42,43\}$; $\{43,44\}$; $\{45,46\}$;
    $\{47,48\}$; $\{48,49\}$; $\{50,51\}$; $\{52,53\}$; $\{53,54\}$;
    $\{55,56\}$; $\{57,58\}$; $\{59,60\}$; $\{60,61\}$; $\{62,63\}$;
    $\{64,65\}$; $\{66,67\}$; $\{67,68\}$; $\{69,70\}$; $\{71,72\}$;
    $\{72,73\}$; $\{74,75\}$; $\{76,77\}$; $\{78,79\}$; $\{80,81\}$;
    $\{81,82\}$; $\{83,84\}$; $\{85,86\}$; $\{86,87\}$; $\{88,89\}$;
    $\{90,91\}$; $\{91,92\}$; $\{93,94\}$; $\{95,96\}$; $\{97,98\}$;
    $\{98,99\}$; $\{100,101\}$; $\{102,103\}$; $\{104,105\}$;
    $\{105,106\}$; $\{107,108\}$; $\{109,110\}$; $\{110,111\}$;
    $\{112,113\}$; $\{114,115\}$; $\{116,117\}$; $\{118,119\}$;
    $\{119,120\}$; $\{121,122\}$; $\{122,138\}$; $\{123,124\}$;
    $\{124,125\}$; $\{126,127\}$; $\{127,128\}$; $\{129,130\}$;
    $\{131,132\}$; $\{133,134\}$; $\{134,135\}$; $\{136,137\}$;
    $\{139,140\}$; $\{141,142\}$; $\{143,144\}$
    \\[2pt]

    12 & A &
    $\{1{:}12\}$;
    $\{5{:}7,13{:}21\}$;
    $\{16{:}18,22{:}30\}$;
    $\{9{:}11,31{:}39\}$;
    $\{7{:}9,20,21,31,32,40{:}44\}$;
    $\{18{:}20,29,30,40,41,45{:}49\}$;
    $\{32{:}34,43,44,50{:}56\}$;
    $\{41{:}43,48{:}51,57{:}61\}$;
    $\{46{:}48,57,58,62{:}68\}$;
    $\{53{:}55,69{:}77\}$;
    $\{51{:}53,60,61,69,70,78{:}82\}$;
    $\{58{:}60,67,68,78,79,83{:}87\}$;
    $\{70{:}72,81,82,88{:}94\}$;
    $\{79{:}81,86{:}89,95{:}99\}$;
    $\{84{:}86,95,96,100{:}106\}$;
    $\{91{:}93,107{:}115\}$;
    $\{89{:}91,98,99,107,108,116{:}120\}$;
    $\{96{:}98,105,106,116,117,121{:}125\}$;
    $\{108{:}110,119,120,126{:}132\}$;
    $\{117{:}119,124{:}127,133{:}137\}$;
    $\{122{:}124,133,134,138{:}144\}$
    \\[2pt]

    30 & A &
    $\{1{:}30\}$;
    $\{7{:}11,18{:}21,29{:}49\}$;
    $\{32{:}34,41{:}44,46{:}68\}$;
    $\{51{:}55,58{:}61,67{:}87\}$;
    $\{70{:}72,79{:}82,84{:}106\}$;
    $\{89{:}93,96{:}99,105{:}125\}$;
    $\{108{:}110,117{:}120,122{:}144\}$
    \\[2pt]

    49 & A &
    $\{1{:}49\}$;
    $\{7{:}11,18{:}21,29{:}68\}$;
    $\{51{:}55,58{:}61,67{:}106\}$;
    $\{89{:}93,96{:}99,105{:}144\}$
    \\[2pt]

    82 & A &
    $\{1{:}82\}$;
    $\{51{:}53,58{:}61,67{:}74,78{:}144\}$
    \\[2pt]

    144 & A &
    $\{1{:}144\}$
    \\

    \end{tabular}
    \end{ruledtabular}
\end{table*}

\begin{figure*}[t]
    \centering
    \includegraphics[width=0.9\linewidth]{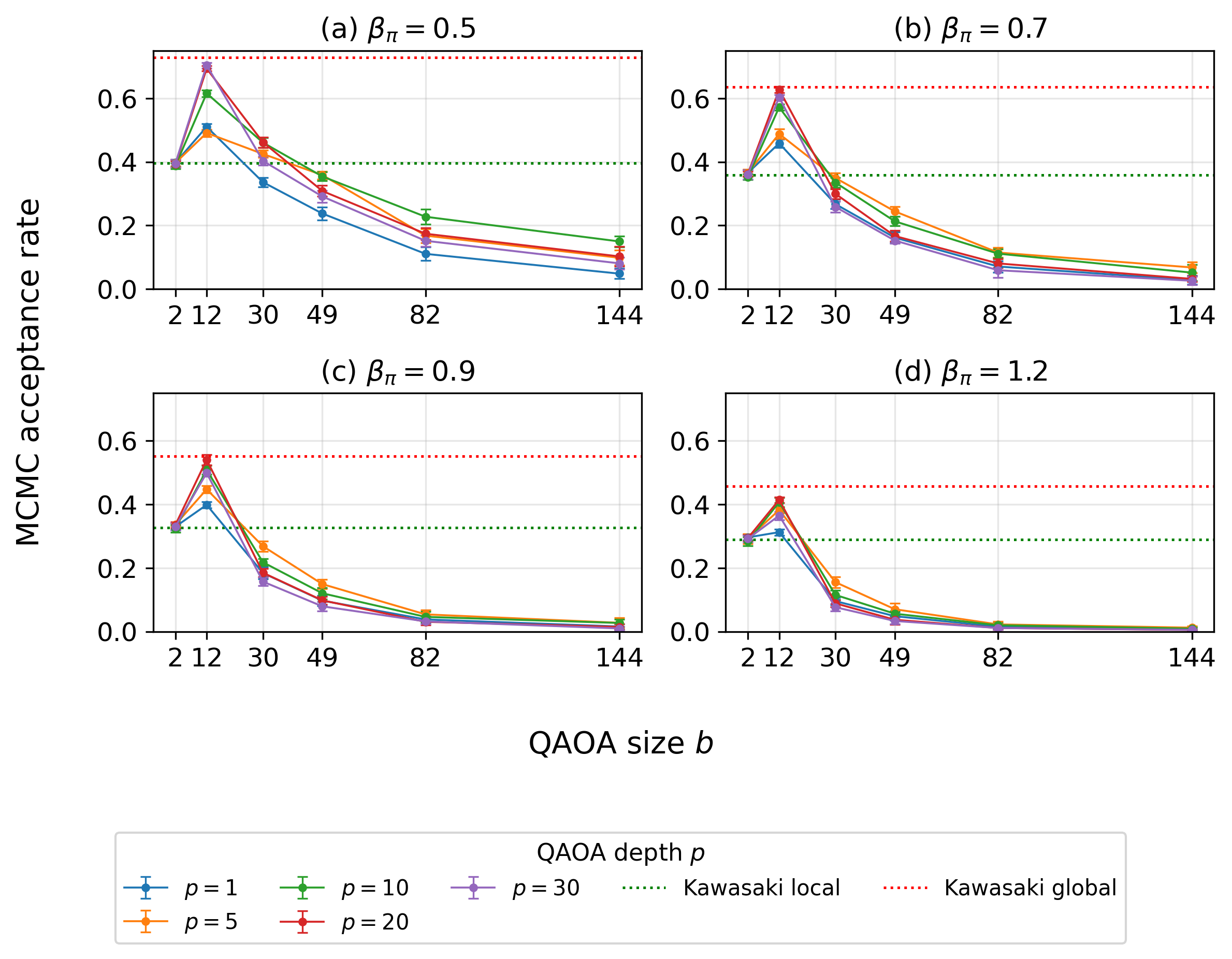}
    \caption{
    MCMC acceptance rates as a function of the block size $b$ for different QAOA depths $p$.
    The dotted green and red lines indicate the local and global Kawasaki baselines, respectively.
    These results supplement the autocorrelation and constraint-leakage results presented in Sec.~\ref{subsec:qaoa-made-results}.
    }
    \label{fig:mcmc-acceptance}
\end{figure*}

Figure~\ref{fig:mcmc-acceptance} presents supplementary MCMC acceptance-rate results for the experiments in Sec.~\ref{subsec:qaoa-made-results}.
The neural-surrogate proposals are compared with global and local Kawasaki dynamics for different block sizes $b$ and QAOA depths $p$.
The reduction in acceptance at larger block sizes accompanies the deterioration in mixing discussed in the main text.
These acceptance rates alone, however, do not identify the causes of this deterioration.

\bibliographystyle{apsrev4-2}
\bibliography{references}

@inproceedings{shor1994algorithms,
  author    = {Shor, Peter W.},
  title     = {Algorithms for Quantum Computation: Discrete Logarithms and Factoring},
  booktitle = {Proceedings of the 35th Annual Symposium on Foundations of Computer Science},
  pages     = {124--134},
  year      = {1994},
  publisher = {IEEE},
  doi       = {10.1109/SFCS.1994.365700}
}

@article{grover1997quantum,
  author  = {Grover, Lov K.},
  title   = {Quantum Mechanics Helps in Searching for a Needle in a Haystack},
  journal = {Physical Review Letters},
  volume  = {79},
  pages   = {325--328},
  year    = {1997},
  doi     = {10.1103/PhysRevLett.79.325}
}

@article{abrams1999quantum,
  author  = {Abrams, Daniel S. and Lloyd, Seth},
  title   = {Quantum Algorithm Providing Exponential Speed Increase for Finding Eigenvalues and Eigenvectors},
  journal = {Physical Review Letters},
  volume  = {83},
  pages   = {5162--5165},
  year    = {1999},
  doi     = {10.1103/PhysRevLett.83.5162}
}

@article{harrow2009quantum,
  author  = {Harrow, Aram W. and Hassidim, Avinatan and Lloyd, Seth},
  title   = {Quantum Algorithm for Linear Systems of Equations},
  journal = {Physical Review Letters},
  volume  = {103},
  pages   = {150502},
  year    = {2009},
  doi     = {10.1103/PhysRevLett.103.150502}
}

@article{preskill2018quantum,
  author  = {Preskill, John},
  title   = {Quantum Computing in the {NISQ} Era and Beyond},
  journal = {Quantum},
  volume  = {2},
  pages   = {79},
  year    = {2018},
  doi     = {10.22331/q-2018-08-06-79}
}

@article{arute2019quantum,
  author  = {Arute, Frank and others},
  title   = {Quantum Supremacy Using a Programmable Superconducting Processor},
  journal = {Nature},
  volume  = {574},
  pages   = {505--510},
  year    = {2019},
  doi     = {10.1038/s41586-019-1666-5}
}

@article{zhong2020quantum,
  author  = {Zhong, Han-Sen and others},
  title   = {Quantum Computational Advantage Using Photons},
  journal = {Science},
  volume  = {370},
  pages   = {1460--1463},
  year    = {2020},
  doi     = {10.1126/science.abe8770}
}

@article{madsen2022quantum,
  author  = {Madsen, Lars S. and others},
  title   = {Quantum Computational Advantage with a Programmable Photonic Processor},
  journal = {Nature},
  volume  = {606},
  pages   = {75--81},
  year    = {2022},
  doi     = {10.1038/s41586-022-04725-x}
}

@article{cerezo2021variational,
  author  = {Cerezo, M. and others},
  title   = {Variational Quantum Algorithms},
  journal = {Nature Reviews Physics},
  volume  = {3},
  pages   = {625--644},
  year    = {2021},
  doi     = {10.1038/s42254-021-00348-9}
}

@article{farhi2014quantum,
  author        = {Farhi, Edward and Goldstone, Jeffrey and Gutmann, Sam},
  title         = {A Quantum Approximate Optimization Algorithm},
  journal       = {arXiv preprint arXiv:1411.4028},
  year          = {2014},
  eprint        = {1411.4028},
  archivePrefix = {arXiv},
  primaryClass  = {quant-ph}
}

@article{gonthier2022measurements,
  author  = {Gonthier, J{\'e}r{\^o}me F. and others},
  title   = {Measurements as a Roadblock to Near-Term Practical Quantum Advantage in Chemistry: Resource Analysis},
  journal = {Physical Review Research},
  volume  = {4},
  pages   = {033154},
  year    = {2022},
  doi     = {10.1103/PhysRevResearch.4.033154}
}

@article{mcclean2018barren,
  author  = {McClean, Jarrod R. and Boixo, Sergio and Smelyanskiy, Vadim N. and
             Babbush, Ryan and Neven, Hartmut},
  title   = {Barren Plateaus in Quantum Neural Network Training Landscapes},
  journal = {Nature Communications},
  volume  = {9},
  pages   = {4812},
  year    = {2018},
  doi     = {10.1038/s41467-018-07090-4}
}

@article{layden2023quantum,
  author  = {Layden, David and Mazzola, Guglielmo and Mishmash, Ryan V. and
             Motta, Mario and Wocjan, Pawel and Kim, Jin-Sung and Sheldon, Sarah},
  title   = {Quantum-Enhanced Markov Chain Monte Carlo},
  journal = {Nature},
  volume  = {619},
  pages   = {282--287},
  year    = {2023},
  doi     = {10.1038/s41586-023-06095-4}
}

@article{nakano2024markov,
  author  = {Nakano, Yuichiro and Hakoshima, Hideaki and Mitarai, Kosuke and Fujii, Keisuke},
  title   = {Markov-Chain Monte Carlo Method Enhanced by a Quantum Alternating Operator Ansatz},
  journal = {Physical Review Research},
  volume  = {6},
  pages   = {033105},
  year    = {2024},
  doi     = {10.1103/PhysRevResearch.6.033105}
}

@article{diez2023quantum,
  author  = {D{\'i}ez-Valle, Pablo and Porras, Diego and Garc{\'i}a-Ripoll, Juan Jos{\'e}},
  title   = {Quantum Approximate Optimization Algorithm Pseudo-Boltzmann States},
  journal = {Physical Review Letters},
  volume  = {130},
  pages   = {050601},
  year    = {2023},
  doi     = {10.1103/PhysRevLett.130.050601}
}

@article{lotshaw2023approximate,
  author  = {Lotshaw, Phillip C. and Siopsis, George and Ostrowski, James and
             Herrman, Rebekah and Alam, Rizwanul and Powers, Sarah and Humble, Travis S.},
  title   = {Approximate Boltzmann Distributions in Quantum Approximate Optimization},
  journal = {Physical Review A},
  volume  = {108},
  pages   = {042411},
  year    = {2023},
  doi     = {10.1103/PhysRevA.108.042411}
}

@inproceedings{germain2015made,
  author    = {Germain, Mathieu and Gregor, Karol and Murray, Iain and Larochelle, Hugo},
  title     = {{MADE}: Masked Autoencoder for Distribution Estimation},
  booktitle = {Proceedings of the 32nd International Conference on Machine Learning},
  series    = {Proceedings of Machine Learning Research},
  volume    = {37},
  pages     = {881--889},
  year      = {2015},
  publisher = {PMLR}
}

@article{uria2016neural,
  author  = {Uria, Benigno and C{\^o}t{\'e}, Marc-Alexandre and Gregor, Karol and
             Murray, Iain and Larochelle, Hugo},
  title   = {Neural Autoregressive Distribution Estimation},
  journal = {Journal of Machine Learning Research},
  volume  = {17},
  number  = {205},
  pages   = {1--37},
  year    = {2016}
}

@inproceedings{larochelle2011neural,
  author    = {Larochelle, Hugo and Murray, Iain},
  title     = {The Neural Autoregressive Distribution Estimator},
  booktitle = {Proceedings of the Fourteenth International Conference on
               Artificial Intelligence and Statistics},
  series    = {Proceedings of Machine Learning Research},
  volume    = {15},
  pages     = {29--37},
  year      = {2011},
  publisher = {PMLR}
}

@article{nakano2025neural,
  author  = {Nakano, Yuichiro and Okada, Ken N. and Fujii, Keisuke},
  title   = {Neural-Network-Assisted Monte Carlo Sampling Trained by
             Quantum Approximate Optimization Algorithm},
  journal = {PRX Quantum},
  volume  = {7},
  pages   = {010338},
  year    = {2026},
  doi     = {10.1103/9nhx-5pym}
}

@article{kawamata2026divide,
  author        = {Kawamata, Yuya and Nakano, Yuichiro and Fujii, Keisuke},
  title         = {Divide-and-Conquer Neural Network Surrogates for Quantum Sampling:
                   Accelerating Markov Chain Monte Carlo in Large-Scale
                   Constrained Optimization Problems},
  journal       = {arXiv preprint arXiv:2604.20701},
  year          = {2026},
  eprint        = {2604.20701},
  archivePrefix = {arXiv},
  primaryClass  = {quant-ph}
}

@article{wang2020xy,
  author  = {Wang, Zhihui and Rubin, Nicholas C. and Dominy, Jason M. and Rieffel, Eleanor G.},
  title   = {{$XY$} Mixers: Analytical and Numerical Results for the
             Quantum Alternating Operator Ansatz},
  journal = {Physical Review A},
  volume  = {101},
  pages   = {012320},
  year    = {2020},
  doi     = {10.1103/PhysRevA.101.012320}
}

@article{montanez2025evaluating,
  author        = {Montanez-Barrera, J. A. and Michielsen, Kristel and
                   Bernal Neira, David E.},
  title         = {Evaluating the Performance of Quantum Process Units
                   at Large Width and Depth},
  journal       = {arXiv preprint arXiv:2502.06471},
  year          = {2025},
  eprint        = {2502.06471},
  archivePrefix = {arXiv},
  primaryClass  = {quant-ph}
}

@article{bucher2026constrained,
  author        = {Bucher, David and Janetschek, Maximilian and Poppel, Michael and
                   Stein, Jonas and Linnhoff-Popien, Claudia and Feld, Sebastian},
  title         = {Constrained Quantum Optimization via Iterative Warm-Start
                   {$XY$}-Mixers},
  journal       = {arXiv preprint arXiv:2604.02083},
  year          = {2026},
  eprint        = {2604.02083},
  archivePrefix = {arXiv},
  primaryClass  = {quant-ph}
}

@article{hadfield2019quantum,
  author  = {Hadfield, Stuart and Wang, Zhihui and O'Gorman, Bryan and
             Rieffel, Eleanor G. and Venturelli, Davide and Biswas, Rupak},
  title   = {From the {Quantum Approximate Optimization Algorithm} to a
             {Quantum Alternating Operator Ansatz}},
  journal = {Algorithms},
  volume  = {12},
  number  = {2},
  pages   = {34},
  year    = {2019},
  doi     = {10.3390/a12020034}
}

@article{montanez2025linear,
  author  = {Montañez-Barrera, J. A. and Michielsen, Kristel},
  title   = {Toward a linear-ramp {QAOA} protocol: evidence of a scaling advantage in solving some combinatorial optimization problems},
  journal = {npj Quantum Information},
  volume  = {11},
  pages   = {131},
  year    = {2025},
  doi     = {10.1038/s41534-025-01082-1}
}

@article{harrigan2021quantum,
  title   = {Quantum approximate optimization of non-planar graph problems on a planar superconducting processor},
  author  = {Harrigan, Matthew P. and Sung, Kevin J. and Neeley, Matthew and Satzinger, Kevin J. and Arute, Frank and Arya, Kunal and Atalaya, Juan and Bardin, Joseph C. and Barends, Rami and Boixo, Sergio and Broughton, Michael and Buckley, Bob B. and Buell, David A. and Burkett, Brian and Bushnell, Nicholas and Chen, Yu and Chen, Zijun and Chiaro, Ben and Collins, Roberto and Courtney, William and others},
  journal = {Nature Physics},
  volume  = {17},
  pages   = {332--336},
  year    = {2021},
  doi     = {10.1038/s41567-020-01105-y}
}

@inproceedings{galda2021transferability,
  author    = {Galda, Alexey and Liu, Xiaoyuan and Lykov, Danylo and
               Alexeev, Yuri and Safro, Ilya},
  title     = {Transferability of Optimal {QAOA} Parameters between Random Graphs},
  booktitle = {2021 IEEE International Conference on Quantum Computing
               and Engineering (QCE)},
  pages     = {171--180},
  year      = {2021},
  doi       = {10.1109/QCE52317.2021.00034}
}

@article{galda2023similarity,
  author  = {Galda, Alexey and Gupta, Eesh and Falla, Jose and Liu, Xiaoyuan and
             Lykov, Danylo and Alexeev, Yuri and Safro, Ilya},
  title   = {Similarity-Based Parameter Transferability in the Quantum
             Approximate Optimization Algorithm},
  journal = {Frontiers in Quantum Science and Technology},
  volume  = {2},
  pages   = {1200975},
  year    = {2023},
  doi     = {10.3389/frqst.2023.1200975}
}

@misc{ibm_riken_system_two,
  author       = {{RIKEN}},
  title        = {{IBM} and {RIKEN} Unveil First {IBM Quantum System Two}
                  Outside of the {U.S.}},
  year         = {2025},
  howpublished = {RIKEN Center for Computational Science},
  url          = {https://www.r-ccs.riken.jp/en/outreach/topics/20250624-1/index.html},
  note         = {Accessed: 2026-08-17}
}

@article{wang2026markovian,
  author        = {Wang, Zeqing and Teske, Julian D. and Bhardwaj, Anshuman and
                   Takahashi, Masahiro O. and Yunoki, Seiji},
  title         = {Simulating the Dynamics of Markovian Quantum Processes
                   by Quantum Collision Models on Quantum Computers},
  journal       = {arXiv preprint arXiv:2606.27856},
  year          = {2026},
  eprint        = {2606.27856},
  archivePrefix = {arXiv},
  primaryClass  = {quant-ph}
}

@article{xu2026filter,
  author        = {Xu, Han and Shirakawa, Tomonori and Yunoki, Seiji},
  title         = {Filter-Assisted Quantum Subspace Diagonalization via
                   Wavefunction Sparsity Engineering},
  journal       = {arXiv preprint arXiv:2605.28040},
  year          = {2026},
  eprint        = {2605.28040},
  archivePrefix = {arXiv},
  primaryClass  = {quant-ph}
}

@article{shinjo2026synchronization,
  author        = {Shinjo, Kazuya and Seki, Kazuhiro and Yunoki, Seiji},
  title         = {Quantum Synchronization and Chimera States in a
                   Programmable Quantum Many-Body System},
  journal       = {arXiv preprint arXiv:2603.11910},
  year          = {2026},
  eprint        = {2603.11910},
  archivePrefix = {arXiv},
  primaryClass  = {quant-ph}
}

@incollection{pelofske2023quantum,
  author    = {Pelofske, Elijah and B{\"a}rtschi, Andreas and Eidenbenz, Stephan},
  title     = {Quantum Annealing vs. {QAOA}: 127 Qubit Higher-Order
               Ising Problems on {NISQ} Computers},
  booktitle = {High Performance Computing},
  series    = {Lecture Notes in Computer Science},
  pages     = {240--258},
  publisher = {Springer},
  year      = {2023},
  doi       = {10.1007/978-3-031-32041-5_13},
  eprint    = {2301.00520},
  archivePrefix = {arXiv}
}

\end{document}